\documentclass[aps,pra,twocolumn]{revtex4-1}

\usepackage{amsmath, amssymb}
\usepackage{bm}
\usepackage{natbib}
\usepackage{color}

\usepackage{graphicx}
\usepackage{float}

\begin{document}

\title{Superradiant Amplification of Acoustic Beams via Medium Rotation}

\author{D. Faccio$^{1,2}$ and E. M. Wright$^{2}$}
\affiliation{$^1$School of Physics \& Astronomy, University of Glasgow, G12 8QQ Glasgow, United Kingdom}
\affiliation{$^2$College of Optical Sciences, University of Arizona, Tucson, Arizona 85721, USA}

\begin{abstract}
Superradiant gain is the process in which waves are amplified via their interaction with a rotating body, examples including evaporation of a spinning black hole and electromagnetic emission from a rotating metal sphere.  {In this paper we elucidate theoretically how superradiance may be realized in the field of acoustics, and predict the possibility of non-reciprocally amplifying or absorbing acoustic beams carrying orbital angular momentum by propagating them through an absorbing medium that is rotating.  We discuss a possible geometry for realizing the superradiant amplification process using existing technology.}
\end{abstract}

\pacs{PACS number(s): }

\date{\today}
\maketitle

\noindent
{\bf{Introduction.}}
Prior to Hawking's epoch-making discovery that black holes must shrink and eventually explode when quantum gravitational effects are accounted for \cite{Hawking}, in 1969 Penrose argued that it must be possible to extract energy from the rotational motion of a spinning black hole \cite{cimento,Nature}. This  idea was then further developed in 1971 by Zel'dovich who argued that spinning black holes must spontaneously evaporate, counter to understanding based on general relativity \cite{Zel1}.   In a tour de force of physical intuition Zel'dovich argued this assertion based on the analogue system of a spinning metallic sphere in which incident vacuum fluctuations of the electromagnetic field, comprised of virtual photons with energy equal to the zero-point value, are converted to real photons, the energy being drawn from the rotational energy of the sphere.  In subsequent work, Zel'dovich and coworkers showed using quantum electrodynamics \cite{Zel2,Zel3} that a metallic cylinder bathed in the electromagnetic vacuum must radiate electromagnetic energy, leading to slowing of the rotation of the cylinder \cite{beken,MIT}.  Zel'dovich argued that what applies to the electromagnetic vacuum fluctuations surrounding a spinning conducting medium must also apply apply to the gravitational vacuum fluctuations, or virtual gravitons, surrounding a black hole, so that spinning black holes must radiate \cite{Zel1}.  Misner  suggested that if real gravitational waves (as opposed to the virtual waves associated with vacuum fluctuations) are incident on a spinning black hole they will be amplified \cite{misner1}, a process that he referred to as superradiance \cite{misner2}. \\
Since these seminal works it has become clear that the complex of ideas surrounding radiation from rotating bodies bathed in the quantum vacuum (now referred to as the Zel'dovich effect), or the related amplification of real waves incident on a rotating absorbing body (now referred to as superradiance), are quite generic as speculated by Zel'dovich and can apply to such disparate wave systems as gravity \cite{misner1,misner2,cardoso}, nonlinear optics \cite{Mar1,Mar2,zeldy}, and matter waves for cold atoms \cite{FedCheSuc06,TakTsuVol08,GhaMus14}.  \\
Our goal in this paper is to elucidate how the process of superradiance may be realized and studied in the field of acoustics, {with the key prediction that it is possible to non-reciprocally amplify or absorb acoustic beams carrying orbital angular momentum by propagating them through an absorbing medium that is rotating.  In this context we note that in comparison to previous treatments of superradiance  in which inward propagating radial waves are amplified upon reflection from the outer surface of a rotating cylinder \cite{Zel1,Zel2,Zel3,beken}, here we consider transmission of an incident beam through the rotating medium.  Our proposal is timely} in view of the recent experimental observation of negative frequencies for acoustic  waves carrying orbital angular momentum that are detected using a pair of rotating microphones \cite{miles}: such negative frequencies signal that the so-called Zel'dovich-Misner condition is satisfied which is a prerequisite for observing superradiant gain \cite{beken}.  \\
{\bf{Governing equations.}}
We consider acoustic waves propagating dominantly along the z-axis in an isotropic and absorbing medium that is rotating around the z-axis at frequency $\Omega$.  Our starting point is the wave equation in the rotating frame for the density variations $\tilde\rho({\bf r},t)$ with respect to the background value $\rho_0>>|\tilde\rho|$ \cite{Rayleigh,Lamb,Boyd}.  Then following the notation of Ref.~\cite{Boyd}, the wave equation in the presence of acoustic absorption is
\begin{equation}\label{start}
{\partial^2 \tilde\rho\over \partial t^2} - \Gamma' \nabla^2 {\partial \tilde\rho\over \partial t} - v^2 \nabla^2  \tilde \rho = 0,
\end{equation}
where $v$ is the velocity of sound, and $\Gamma'$ is the damping parameter. To proceed, we first transfer to the non-rotating laboratory frame in cylindrical coordinates ${\bf r}=(r,z,\phi)$ using the change of variables \cite{Zel1,beken}
\begin{equation}
t'=t, \quad r'=r, \quad z'=z, \quad \phi'=\phi+\Omega t.
\end{equation}
Then, performing a change of variables to the non-rotating (primed) frame, and dropping the primes for simplicity in notation, we obtain
\begin{equation}\label{WE}
\left ( {\partial \over \partial t}  + \Omega {\partial\over \partial \phi }\right )^2\tilde\rho - \Gamma' \nabla^2 \left ( {\partial \over \partial t}  + \Omega {\partial\over \partial \phi }\right ) \tilde\rho - v^2 \nabla^2  \tilde \rho= 0.
\end{equation}
This wave equation in the laboratory frame is the basis for the following analysis.\\

\begin{figure}[]
\includegraphics[width=0.5 \textwidth]{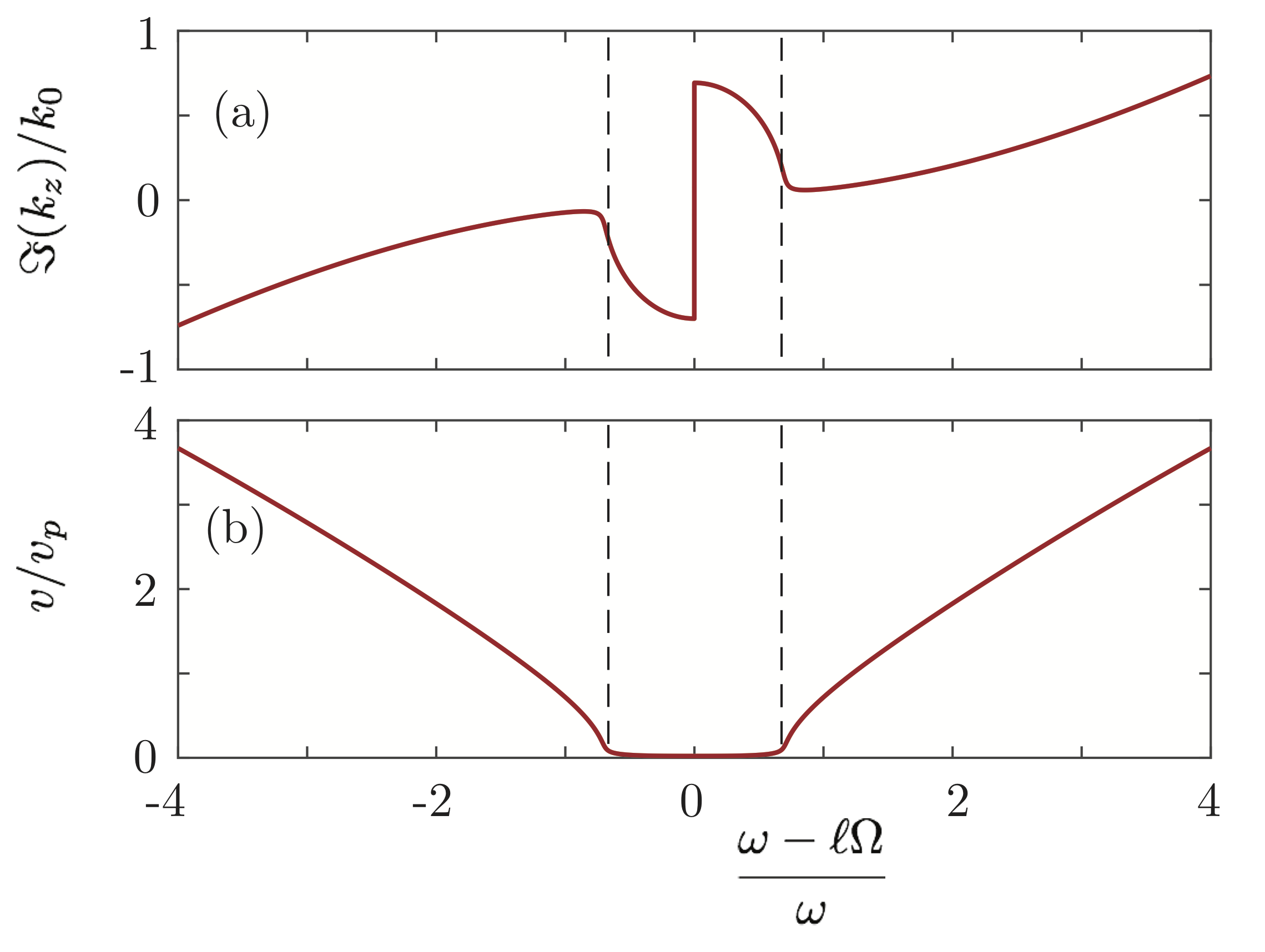}
\caption{\label{FigR1}   Plots of (a) the absorption ${\Im(k_z)/ k_0}$ and (b) the scaled {\it inverse} phase velocity ${v/ v_p}={\Re(k_z)/ k_0}$ both as functions of ${(\omega-\ell\Omega)/ \omega}$, and for parameters $\eta=0.1$ and ${k_r/k_0}=0.7$.  The vertical dashed lines delineate the range in Eq. (\ref{range}). Negative ${\Im(k_z)/ k_0}$ implies gain.}
\end{figure}

\begin{figure}[]
\includegraphics[width=0.5 \textwidth]{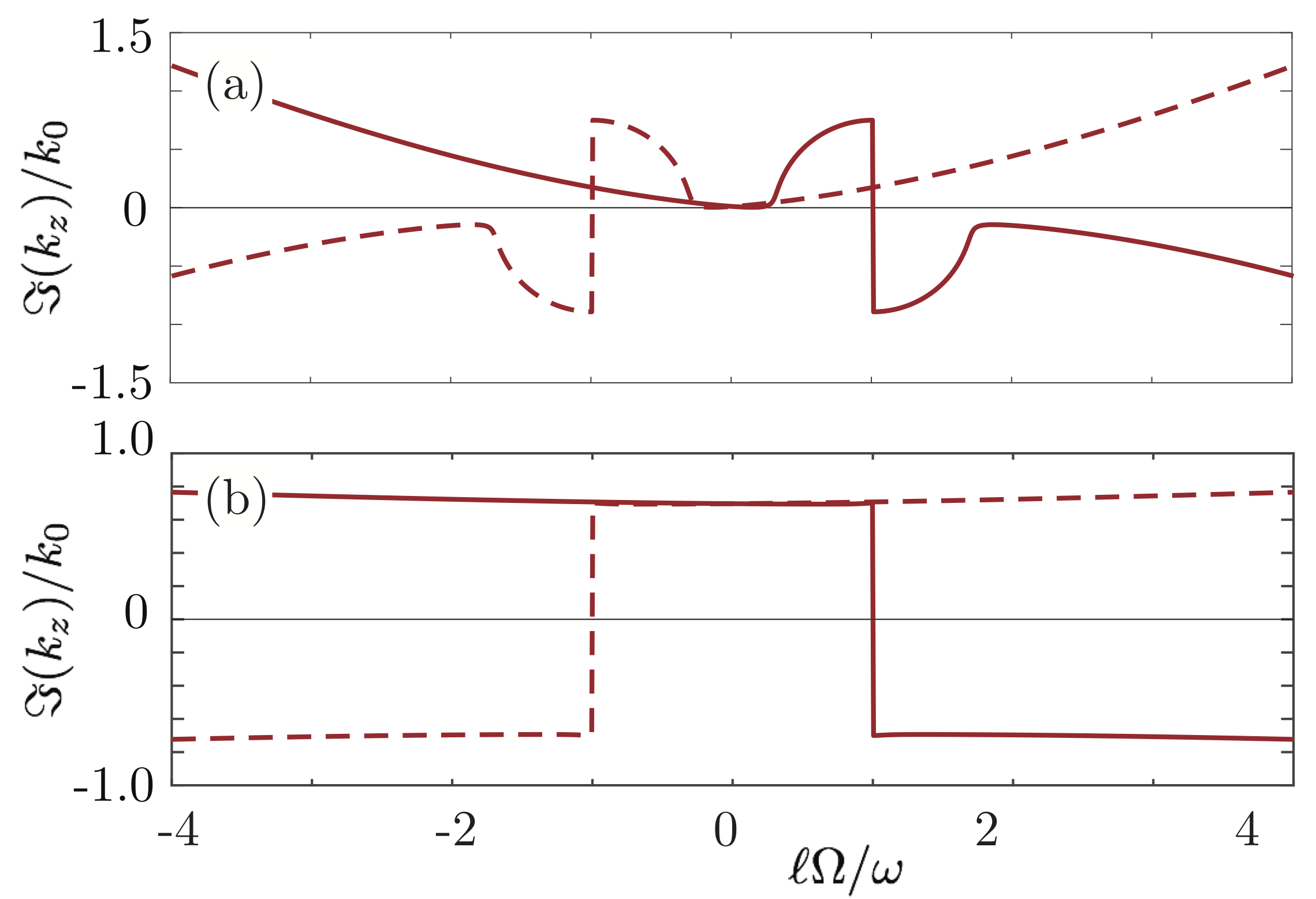}
\caption{\label{FigNonRec} Scaled absorption ${\Im(k_z)/ k_0}$ versus ${\ell\Omega/\omega}$ for the forward (solid line) and backward (dashed line) propagating Bessel beam solutions, and parameters ${k_r/ k_0}=0.7$ (chosen to match the conditions of Ref.~\cite{miles}) and (a) {$\eta=0.1$; (b) $\eta=10$.}}
\end{figure}

\noindent{\bf{Bessel beams.}}
Our analysis is based on the fact that Eq.~(\ref{WE}) has exact Bessel beam solutions that carry orbital angular momentum (OAM) \cite{BarThoMar16,JimSanRom16}.  In particular we consider an acoustic wave of winding number $\ell$ and frequency $\omega>0$, and in complex representation the exact solution is
\begin{equation}\label{BB}
\tilde\rho(r,z,\phi,t) = AJ_{|\ell|}(k_r r)e^{i(\ell\phi-\omega t + k_z z) }  .
\end{equation}
Here $k_r$ is the radial component of the acoustic wave vector and $k_z$ the z-component.  We choose $k_z$ (see below) such that its real part $\Re (k_z)$ is greater than zero, in which case the solution Eq.~(\ref{BB}) represents a Bessel beam propagating along the positive z-axis.\\
{\bf{Dispersion relation.}}
Substituting the exact solution (\ref{BB}) into the acoustic wave equation Eq.~(\ref{WE}), in the laboratory frame we obtain the exact dispersion relation
\begin{equation}\label{disprel}
k_r^2 + k_z^2 = \frac{(\omega-\ell\Omega)^2/v^2}{1-i\Gamma'(\omega-\ell\Omega)/v^2} ,
\end{equation}
the solution of which may be written as
\begin{equation}\label{disprel2}
{k_z\over k_0}  = \pm\sqrt{
\frac{[(\omega-\ell\Omega)/\omega]^2}{1-i\eta[(\omega-\ell\Omega)/\omega]}  - {k_r^2\over k_0^2} } ,
\end{equation}
where $\eta=\Gamma'\omega/v^2$ is a positive dimensionless parameter characterizing the acoustic absorption per length $1/k_0$, and $k_0=\omega/v$.  To use Eq.~(\ref{disprel2}), we choose a value for $k_r<k_0$, and Eq.~(\ref{disprel}) is then solved for $k_z$: as stated above, we choose the sign in the solution such that $\Re (k_z)>0$.  The absorption experienced by the acoustic field is proportional to the imaginary part $\Im(k_z)$, and as a measure of this we consider the absorption per length $1/k_0$:
\begin{equation}
{\alpha\over k_0} = {\Im(k_z)\over k_0}  .
\end{equation}
Another key quantity is the phase-velocity $v_p={\omega/ \Re(k_z)}$, and it is useful to consider the phase-velocity scaled to $v=\omega/k_0$,
\begin{equation}
\frac{v_p}{v} = {k_0\over\Re(k_z)}  .
\end{equation}
A supersonic phase-velocity arises if ${v_p/ v}>1$, and a subsonic phase-velocity if ${v_p/ v}<1$.\\
We now examine the dispersion relation searching for conditions under which superradiant amplification can arise for individual Bessel beams. Since a general input beam can be viewed as a wavepacket of Bessel beams the insights so gained will apply to more general incident acoustic beams.\\
{\bf{Superradiant amplification.}}
As an illustration of the superradiant amplification that can arise, Fig.~\ref{FigR1} shows plots of (a) the scaled absorption ${\Im(k_z)/ k_0}$ and (b) the scaled {\it inverse} phase velocity ${v/ v_p}={\Re(k_z)/ k_0}$, both as functions of ${(\omega-\ell\Omega)/\omega}$, and for parameters $\eta=0.1$ and ${k_r/ k_0}=0.7$.  What is striking from Fig.~\ref{FigR1}(a) is that the scaled absorption is negative for $(\omega-\ell\Omega)<0$ signifying gain, with the transition from absorption to gain occurring in the range
\begin{equation}\label{range}
-{k_r\over k_0}  < {(\omega-\ell\Omega)\over \omega}  <  {k_r\over k_0}  ,
\end{equation}
which is delineated by the vertical dashed lines, and maximum scaled gain $-{\Im(k_z)/ k_0}={kr/ k_0}$ in this range.  Physically, the energy needed for this superradiant amplification  is drawn from the rotational energy of the medium, which means that energy must be supplied to keep the medium rotating at the constant rate $\Omega$ or the medium would spin down. \\
Based on the above analysis, the condition for superradiant amplification is
\begin{equation}\label{SR}
\omega-\ell\Omega <0.
\end{equation}
which coincides with the Zel'dovich-Misner condition for superradiant gain \cite{beken}, and corresponds to when the rotationally shifted frequency of the source in the rotating frame becomes negative.\\
Accompanying the superradiant amplification in the range Eq.~(\ref{range}) we find that the inverse scaled phase-velocity shown in Fig.~\ref{FigR1}(b) is supersonic: the scaled phase-velocity ${v_p/ v}$ is infinite at $(\omega-\ell\Omega)=0$, and significantly larger than $v$ over the rest of the range.  Another way to look at this is to note that $\Re({k_z})=0$ for $(\omega-\ell\Omega)=0$, implying an acoustic impedance, $Z$, that is close to zero in the range in Eq.~\eqref{range}. We can then talk of a $Z$-near-zero (ZNZ) regime, in analogy to epsilon-near-zero (ENZ) physics in optics \cite{engheta} and note that the two regimes share many common aspects that are related  to the nature of wave propagation in the presence of a vanishing impedance.  Both ENZ optics and ZNZ acoustics exhibit a series of signatures such as the aforementioned large or infinite phase velocity, a concomitant lengthening to infinity of the wavelength and a close-to-zero group velocity \cite{boyd_slow}. The ZNZ point at which $(\omega-\ell\Omega)=0$, cannot be accessed due to the extreme consequences of a zero-group velocity (or infinite phase velocity and purely DC, non-oscillatory nature of the wave) that imply a complete reflection of the incoming wave.   Therefore, the discontinuity from absorption to gain at  $(\omega-\ell\Omega)=0$, although not ruled out from thermodynamic considerations \cite{beken}, is not a problem as it is not physically accessible. Nevertheless, in analogy with ENZ studies in optics, we can still couple waves into the medium in the ZNZ regime (away from the $Z=0$ point) and observe enhanced acoustic interactions.\\
We note that significant absorption and gain occur also outside the range given by Eq.~(\ref{range}).  This is illustrated in Fig.~\ref{FigR1}(a) that shows absorption, $(\omega-\ell\Omega)>0$, and gain, $(\omega-\ell\Omega)<0$, outside the range in Eq.~(\ref{range}) and with accompanying phase velocities that can be supersonic or supersonic, see Fig.~\ref{FigR1}(b). \\
\noindent
{\bf{Non-reciprocity.}}
We consider the backward propagating version of Eq. (\ref{BB})
\begin{equation}\label{BBB}
\tilde\rho_b(r,z,\phi,t) = AJ_{|\ell|}(k_r r)e^{i(-\ell\phi-\omega t -k_z z) } .
\end{equation}
where $k_z$ is again chosen such that its real part $\Re (k_z)$ is greater than zero, in which case the solution (\ref{BBB}) represents a Bessel beam propagating along the negative z-axis.  Here $+\ell$ has been replaced by $-\ell$ so that sense of rotation of the OAM is preserved with the change of propagation direction.  Then for propagation along the negative z-axis the corresponding solution of the dispersion relation is
\begin{equation}\label{disprel2B}
{k_z\over k_0}  = \sqrt{
\frac{[(\omega+\ell\Omega)/\omega]^2}{1-i\eta[(\omega+\ell\Omega)/\omega]}  - {k_r^2\over k_0^2} } .
\end{equation}
Since the form of $k_z$ for the counter-propagating waves in Eqs.~(\ref{disprel2}) and (\ref{disprel2B}) differ by the factors $(\omega\pm\ell\Omega)$ this means that for a fixed rotation rate counter-propagating acoustic beams with the same OAM can experience different gain or absorption.  That is, there is non-reciprocity with respect to propagation direction that can be used to realize sound isolation \cite{Fleury,Yang}.\\
An example of this non-reciprocity is shown in Fig.~\ref{FigNonRec}(a) where the scaled absorption ${\Im(k_z)/ k_0}$ is shown as a function of ${\ell\Omega/\omega}$ for the forward (solid line) and backward (dashed line) propagating Bessel beam solutions, and parameters {$\eta=0.1$ and ${k_r/ k_0}=0.7$}.  This figure has several interesting features: first, if we consider a given propagation direction (solid or dashed line) then for a fixed rotation rate $\Omega$ it is clear that winding numbers $\pm\ell$ experience different absorptions, and this highlights the dichroism for acoustic OAM alluded to earlier.  Second, in the range $\left | {\ell\Omega/\omega} \right | <  1$ both directions of propagation experience absorption but one more than the other: This highlights the non-reciprocity discussed above, and if the differential absorption times length product can be made large enough this can be used to realize sound isolation.  Third, in the complimentary range $\left | {\ell\Omega/\omega} \right | >  1$ one direction of propagation experiences gain while the other experiences absorption.  To the best of our knowledge this is a case never discussed before \cite{Fleury,Yang}.\\

\begin{figure}[]
\includegraphics[width=0.5 \textwidth]{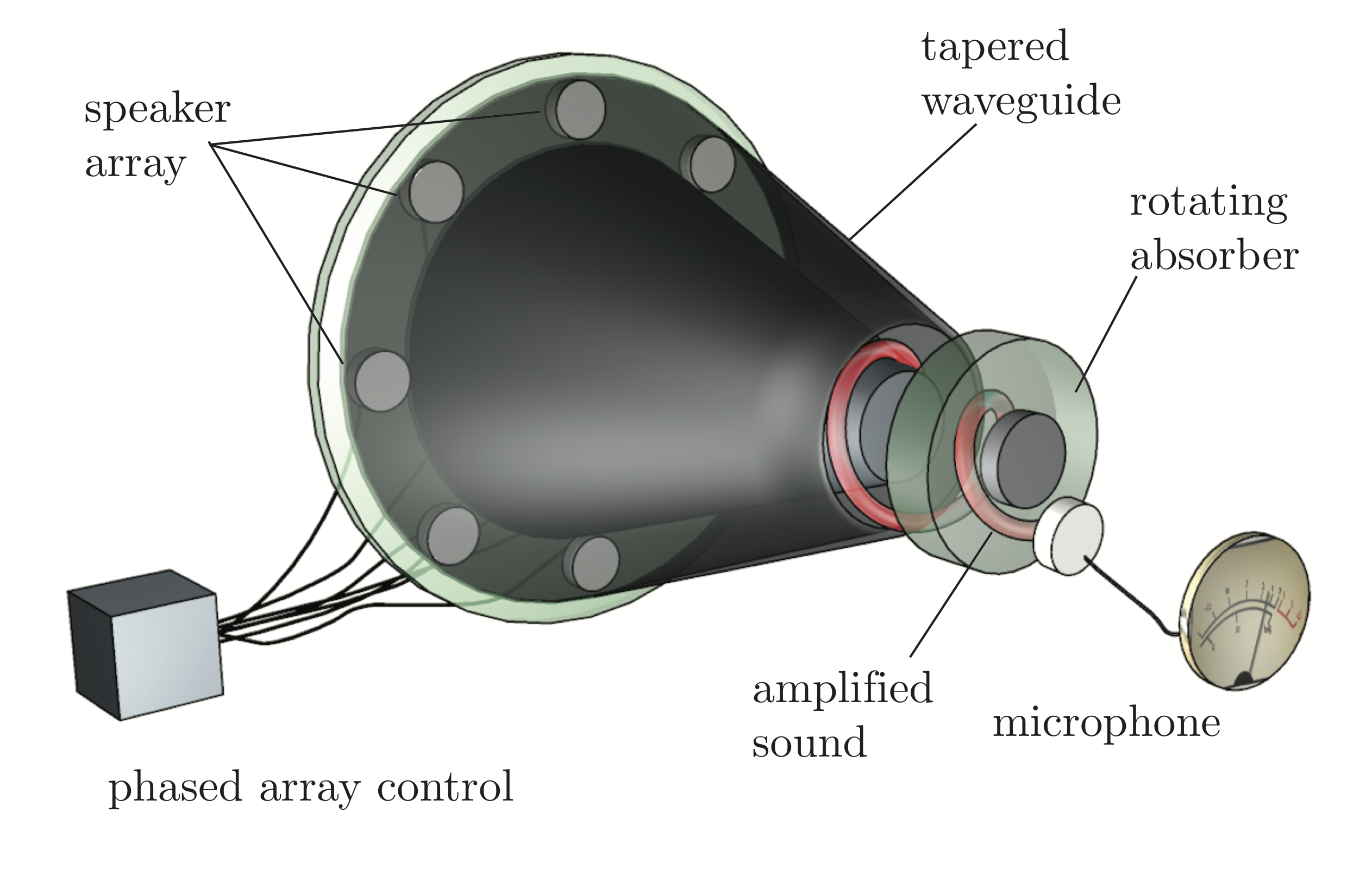}
\caption{\label{exp} Schematical representation of a possible experimental layout that would allow to measure superradiant gain by extraction of energy from the rotation of an absorbing disk.}
\end{figure}

\noindent {\bf{Experimental considerations.}}
One of the key features associated with superradiant gain is the change in sign of the material absorption from positive to negative that occurs when the Rotational-Doppler shifted frequency becomes negative. Such a negative-frequency condition was observed recently in the experiments reported by  Gibson et al. \cite{miles} with a microphone on a platform rotating at $\sim10-50$ Hz and placed in the path of a (non-rotating) acoustic wave source with 90 Hz oscillation frequency (in the laboratory frame) and OAM $\ell=4$. In these experiments, the inversion of the OAM sign, or temporal inversion of the beam occurs when $\omega-\ell\Omega<0$, thus clearly setting the scene for observing superradiant gain given that the key condition Eq.~(\ref{SR}) can be satisfied in experiments. However, in order to observe gain one must extend the experiment by adding absorption: this can be simply achieved by placing an absorbing layer in front of the microphones. In Fig.~\ref{exp} we show an example of a possible experimental layout that would satisfy the conditions used in our analysis. Generation of the acoustic OAM beam follows the same scheme as in \cite{miles}: the speakers form a phase array of sine-wave emitters. By increasing the acoustic sine-wave phase from one speaker to the next, full phase shifts of multiples $\ell$ of $2\pi$ will give rise to acoustic OAM beams that are then directed towards the rotating absorbing disk via a hollow waveguide. A microphone is placed behind this disk that measures the transmitted sound intensity. The signature of superradiant gain would be a crossover from absorption to gain as the rotation speed is increased and Eq.~(\ref{SR}) is satisfied. We note that gain or phrased differently, an increase in phonon number will not depend on the reference frame in which it is measured. It is therefore possible to either use microphones that are stationary in the lab frame or co-rotating solidly with the disk, as in \cite{miles}. This invariance of the gain with respect to the measurement reference frame can be directly verified starting from Eq.~(\ref{start}) and performing the analysis in the co-rotating frame.   \\
It also worth noting that the acoustic wavelength will typically be an order of magnitude or more greater than the dimensions of the overall experiment and more than two orders of magnitude larger than than the absorbing disk thickness. The waveguide structure therefore guides the wave towards the absorbing disk \cite{miles} where, in this deeply sub-wavelength regime, a maximum of 50\% absorption can be expected for a stationary medium \cite{CPA1,CPA2}. Standard acoustic absorbing foam will typically absorb $\sim10$\% at $100$ Hz  (for a 1 cm thick slab) and would correspond to the case $\eta\sim10$, see Fig.~\ref{FigNonRec}(b). From {Fig.~\ref{FigNonRec}}(b) one can infer that existing technology is predicted to give rise to an acoustic gain of order 1-10\% and should therefore be readily observable.\\
\noindent {\bf{Conclusions.}}
Superradiant amplification in the form of amplification from a spinning absorber is very generic and was first discovered in the context of spinning black holes. Yet, it has not been observed to date due to the difficulty in satisfying the required conditions, in particular mechanical rotation frequencies that are close to the wave frequencies. Here we provide details of an acoustic system that could realistically realise such conditions with currently available experimental setups. Moreover, our analysis has unveiled  new physical effects, i.e. non-reciprocal absorption and gain and an enhanced superradiant gain that occurs within a specific window of mechanical rotation frequencies, Eq.~(\ref{range}), corresponding to the $Z$-near-zero, ZNZ, regime. In this region the group velocity is close to zero and the enhanced gain can be ascribed to ``slow wave'' effects, i.e. to an increase in the mode density, which enhances interaction with the medium, in analogy with slow light absorption and gain in optics \cite{slow1,slow2}. We expect experiments in the near future to provide the first evidence of superradiance i.e. the increase of acoustic wave energy at the expense of mechanical rotational energy of an absorbing medium.  \\
{\bf{Acknowledgements.}}
The authors thank Profs. Miroslav Kolesik, Masud Mansuripur of the College of Optical Sciences at the University of Arizona and Ermes Toninelli, Miles Padgett at the University of Glasgow for useful discussions and suggestions regarding this work. DF acknowledges financial support from EPSRC (UK, Grant No. EP/P006078/2).


\begin{thebibliography}{99}

\bibitem{Hawking} S.W. Hawking, Nature {\bf{248}}, 30  (1974).
 \bibitem{cimento} R. Penrose, Riv. Nuovo Cimento, Numero Speziale I, 257 (1969), reprinted in Gen. Rel. Grav., {\bf 34} 1141 (2002) (1969).
 \bibitem{Nature} R. Penrose and B. M. Floyd, Nature {\bf{ 229}},1 77 (1971).
 
\bibitem{Zel1} Y. B. Zel'dovich, Pis'ma Zh. Eksp. Teor. Fiz. 14, 270 (1971), translated in JETP Lett. 14, 180 (1971).
 \bibitem{Zel2} Y. B. Zel'dovich, Zh. Eksp. Teor. Fiz. 62, 2076 (1972), translated in Sov. Phys. JETP 35, 1085 (1972).
 \bibitem{Zel3} Ya. B. Zel'dovich, L. V. Rozhanskii, A. A. Starobinskii, Izvestiya Vysshikh Uchebnykh Zavedenii, Radiofizika, {\bf 29}, I008-I016 (1986). 
 \bibitem{beken} J. D. Bekenstein, M. Schiffer, Phys. Rev. A  {\bf 58}, 064014 (1998).
 \bibitem{MIT} M. F. Maghrebi, R. L. Jaffe, M. Kardar, Phys. Rev. Lett. {\bf 108}, 230403 (2012).
 \bibitem{misner1} C.W. Misner, Phys. Rev. Lett. {\bf 28}, 994 (1972).
\bibitem{misner2} C. W. Misner, Bull. Amer. Phys. Soc. {\bf 17}, 472 (1972).
\bibitem{cardoso}  R. Brito,  V. Cardoso, and P. Pani, Lecture Notes in Physics 906, {\emph{Superradiance}}, Springer (2015).
\bibitem{Mar1} F. Marino, Phys. Rev. A {\bf 78}, 063804 (2008).
\bibitem{Mar2} F. Marino, M. Ciszak, and A. Ortolan, Phys. Rev. A {\bf 80}, 065802 (2009). 
\bibitem{zeldy} D. Faccio, E.M. Wright, Phys. Rev. Lett. {\bf 118}, 093901 (2017).
\bibitem{FedCheSuc06} F. Federici, C. Cherubini, S. Succi, and M. Tosi, {\sl Phys.Rev. A} {\bf 73} 033604 (2006).
\bibitem{TakTsuVol08} H. Takeuchi, M. Tsubota, and G. Volovik, { J. Low. Temp. Phys.} {\bf 150}, 624 (2008).
\bibitem{GhaMus14} N. Ghazanfari and O. E. Mustecaplioglu, {Phys.Rev. A} {\bf 89}, 043619 (2014).
\bibitem{miles}  G. M. Gibson {\it et al.}, Proc. Nation. Acad. Sci. (2018).  
\bibitem{Rayleigh} J. W. S. Rayleigh, {\it Theory of Sound} (Dover, New York, 1896/1945), Vol. 2.
\bibitem{Lamb} Lamb, H., 1932, Hydrodynamics (Cambridge University Press, Cambridge).
\bibitem{Boyd} R. W. Boyd {\it Nonlinear Optics} $3^{rd}$ Edition, (Academic Press, Amsterdam, 2008) Chap. 8.
\bibitem{BarThoMar16} D. Baresch, J.-L. Thomas, and R. Marchiano, Phys. Rev. Lett. 116 (2016).
\bibitem{JimSanRom16} N. Jimenez, R. Pico, V. Sanchez-Morcillo, V. Romero-Garc'a, L. M. Garc'a-Raffi, and K. Staliunas, Phys. Rev. E 94, 053004 (2016).
\bibitem{engheta} I. Liberal, N. Engheta, Nat. Photon. {\bf{11}}, 149 (2017).
\bibitem{boyd_slow} K.L. Tsakmakidis, O. Hess, R.W. Boyd, Z. Zhang, Science {\bf 358}, 319 (2017).
\bibitem{Fleury} R. Fleury, D. L. Sounas, C. F. Siek, M. R. Haberman, and A. Alu, {Science}  {\bf 343}, 516 (2014).
\bibitem{Yang} Z. Yang, F. Gao, X. Shi, X. Lin, Z. Gao, Y. Chong, and B. Zhang, { Phys. Rev. Lett.} {\bf 114}, 114301 (2015).
\bibitem{CPA1} S. Thongrattanasiri,  F. H. L. Koppens, F. J. Garcia de Abajo,   Phys. Rev. Lett. {\bf 108}, 047401 (2012).
\bibitem{CPA2} T. Roger {\it et al.}, { Nature Commun.} {\bf 6}, 7031 (2015).
\bibitem{slow1} L. O'Faolain, et al.  Opt. Express {\bf 18}, 27627 (2010).
\bibitem{slow2} S. Ek, P. Lunnemann, Y. Chen, E. Semenova, K. Yvind, J. Mork, Nature Commun. {\bf 5}, 5039 (2014)

\end{thebibliography}
\end{document}